\documentclass[11pt]{amsart}
\usepackage[T1]{fontenc}
\usepackage[margin=1.5in]{geometry}
\usepackage{amssymb, amsmath, amsthm}
\usepackage{graphicx, tikz}
\usepackage{txfonts}

\usepackage{xcolor}
\usepackage{enumerate}
\numberwithin{equation}{section}
\usepackage[textsize=tiny]{todonotes}

\DeclareMathOperator*{\argmin}{argmin\ }

\let\oldtocsection=\tocsection
\let\oldtocsubsection=\tocsubsection
\let\oldtocsubsubsection=\tocsubsubsection
\renewcommand{\tocsection}[2]{\hspace{0em}\oldtocsection{#1}{#2}}
\renewcommand{\tocsubsection}[2]{\hspace{1em}\oldtocsubsection{#1}{#2}}
\renewcommand{\tocsubsubsection}[2]{\hspace{2em}\oldtocsubsubsection{#1}{#2}}

\definecolor{ocean}{rgb}{0,0.1,0.6}

\definecolor{imperialGreen}{RGB}{2,137,59}
\definecolor{imperialBlue}{RGB}{0, 62, 116}
\definecolor{imperialBrick}{RGB}{165,25,0}
\definecolor{imperialProcess}{RGB}{0,133,202}

\usepackage{hyperref}
\hypersetup{colorlinks=false, linkbordercolor=imperialProcess,
citebordercolor=imperialBrick}

\newcommand{\ind}{1\hspace{-2.1mm}{1}}
\newcommand{\eps}{\varepsilon}
\newcommand{\half}{\frac{1}{2}}

\newcommand{\D}{\mathrm{d}}
\newcommand{\EE}{\mathbb{E}}
\newcommand{\pf}{\mathfrak{p}}
\newcommand{\Kk}{\mathcal{K}}
\newcommand{\Pp}{\mathcal{P}}
\newcommand{\Ppb}{\overline{\mathcal{P}}}
\newcommand{\Pb}{\boldsymbol{P}}
\newcommand{\Tt}{\mathcal{T}}
\newcommand{\Ll}{\mathcal{L}}

\newcommand{\sigimp}{\sigma_{\mathrm{imp}}}
\newcommand{\signn}{\sigma_{\mathrm{NN}}}

\theoremstyle{definition}

\makeatletter
\newcommand{\subjclassjel}[2][1991]{%
  \let\@oldtitle\@title%
  \gdef\@title{\@oldtitle\footnotetext{#1 \emph{JEL subject classification.} #2}}%
}
\graphicspath{{Figures/}}

\title{Deep learning interpretability for rough volatility}

\author{Bo Yuan}
\email{by258@cam.ac.uk}
\address{Judge Business School, University of Cambridge, United Kingdom.}

\author{Damiano Brigo}
\email{damiano.brigo@imperial.ac.uk}
\address{Department of Mathematics, Imperial College, London, United Kingdom.}
\author{Antoine Jacquier}
\email{a.jacquier@imperial.ac.uk}
\address{Department of Mathematics, Imperial College, London, United Kingdom, and The Alan Turing Institute.}
\author{Nicola Pede}
\email{nicola.pede@mediobanca.com}
\address{Mediobanca, Financial Engineering, London SW1E 6AJ, United Kingdom.}

\date{November 28, 2024}
\thanks{\small The opinions here expressed  are solely those of the authors and do not represent in any way those of their employers. 
AJ acknowledges the support of the EPSRC grants EP/T032146/1 and P/W032643/1.  For the purpose of open access, the author(s) has applied a Creative Commons Attribution (CC BY) licence (where permitted by UKRI, ‘Open Government Licence’ or ‘Creative Commons Attribution No-derivatives (CC BY-ND) licence’ may be stated instead) to any Author Accepted Manuscript version arising’.}

\begin{document}
\begin{abstract}
Deep learning methods have become a widespread toolbox for pricing and calibration of financial models.
While they often provide  new directions and research results, their `black box' nature also results in a lack of interpretability.
We provide a detailed interpretability analysis of these methods in the context of rough volatility -- a new class of volatility models for Equity and FX markets. 
Our work sheds light on the neural network learned inverse map between the rough volatility model parameters, seen as mathematical model inputs and network outputs, and the resulting implied volatility across strikes and maturities, seen as mathematical model outputs and network inputs. 
This contributes to building a solid framework for a safer use of neural networks in this context and in quantitative finance more generally.
\end{abstract}

\maketitle

\vspace{-1.5cm}

\tableofcontents

\section{Introduction}
\label{sec:introduction}
The advent of deep learning technology has initiated rapid and deep changes within the financial industry, on the retail side and on both the commercial and investment banking sides, over the last decade. 
For pricing purposes, stochastic volatility models are ubiquitous;
while "classical" stochastic volatility models are by now well understood, their limitations, and in particular their relative inability to replicate the at-the-money skew, have initiated new research in so-called "rough volatility models"~\cite{roughvol}, which provide better historical volatility estimation and better at-the-money skew fits.
With additional complexity obviously come  numerical challenges, and pricing and calibration are not as (relatively) straightforward as for classical stochastic volatility models.
While recent advances~\cite{bayer2020hierarchical, bayer2024roughbook, bennedsen2017hybrid, bonesini2021functional, jacquier2018vix} have made the problem easier, a natural question is to investigate whether new deep-learning-based techniques could also help in this direction.
This investigation, both for pricing and for calibration on diverse financial markets (Equity, Fixed Income), has in fact been carried out extensively over the past few years~\cite{alaya2021deep, baschetti2024deep, bayer2019deep, benth2021accuracy, itkin2019deep, jacquier2023random, liu2019neural, rosenbaum2021deep} with increasing degrees of success.

There exists, however, a fundamental trade-off in model development, between model complexity and model understanding: a simple mathematical model, while not perfect for fitting the market, is sometimes preferred over complex models, as its parameters and limitations (hence the risks it may entail) can be monitored accurately.
On the other hand, complex nonlinear models stemming from machine learning offer distinct advantages, including the  ability to achieve a good fit to market data, better reactions to sudden market changes and the ability to deal with large datasets.
This, however, comes at the cost of a relative loss of risk monitoring, as model parameters' impacts on prices and implied volatilities cannot be clearly disentangled and understood in complex nonlinear models.
In general, deep learning methods intrinsically suffer from this issue, as the high number of parameters of a neural network makes it impossible to accurately measure the impact on the output.
This has led to seeing neural networks (NN) as `black boxes', whereby the map between  NN input parameters and outputs is highly singular. 
"Interpretability" therefore is key to promoting deep learning methods further, as it provides a safety net for risk monitoring.
While a rigorous mathematical formulation of Interpretability is not yet available,
it has nevertheless been characterised informally as~\cite{millerExplanation} \emph{`the extent to which a human can comprehend the cause of a decision'} or~\cite{kim2016examples} \emph{`the extent to which a human can reliably anticipate the model's conclusion'}. In this sense interpretability is on a spectrum, and the idea is that the more one can comprehend a model decision or anticipate the model conclusion, the more the model is interpretable. These definitions are not fully convincing, and they are further discussed in \cite{molnar2020interpretable}. We should also mention that, at times, the terms interpretability and the related term explainability (as opposed to explanation) are used interchangeably, see again \cite{molnar2020interpretable}. We will not use the term ``explainability" in this paper, but only resort to interpretability. 

In this paper, we plan to use deep neural networks as a shortcut to invert efficiently the mathematical model parameters to implied volatility map across strikes and maturities in the rough Heston model~\cite{el2019roughening, el2019characteristic, euch2018perfect}, 
which has become one of the main models for  rough volatility.
We develop an interpretability analysis of deep learning for pricing, 
namely understanding the deep neural network \emph{learned} map between the mathematical model implied volatilites across strikes and maturities seen as inputs and the corresponding mathematical model parameters seen as outputs.
To do so, we first generate a dataset of implied volatilities from several rough Heston models parameter sets, using standard methods for the mathematical model, and subsequently fit this data set with a neural network. Armed with this, we study the  interpretability of this deep learning approach, and see what it says on the impact of the parameters on the volatility patterns when compared with classical intuitions on  the model driven by mathematical and financial considerations.
We only consider Feedforward neural networks (FNN) here, as they seem to outperform their Convolutional relatives, see~\cite{brigo2021interpretability}, where the interpretability of the standard Heston model analogous map is studied, but a similar analysis can easily be carried out to other neural network architectures.

Our interpretability analysis focuses on several interpretability methods recently proposed in the literature:
the Local Interpretable Model-agnostic Explanations~\cite{ribeiro2016should}
(\texttt{LIME}) method focuses on interpreting individual predictions locally;
the \texttt{Gradient*Input} method  
refines standard gradient methods by multiplying the latter by the input feature;
\texttt{DeepLIFT}~\cite{shrikumar2017learning}
improves upon the latter with additional backwards propagation;
Layer-wise Relevance Propagation (\texttt{LRP})~\cite{bach2015pixel} is a breakdown of non-linear classifiers;
Shapley Additive explanations (\texttt{SHAP})~\cite{lundberg2017unified}, originating from cooperative game theory, is a global interpretability method that aims to improve interpretability by estimating the significance values of characteristics for certain predictions. 

The structure of the paper is as follows:
Section~\ref{sec:interpr} introduces  interpretability methods needed for the analysis;
Section~\ref{sec:heston} focuses on the calibration of a feedforward neural network to  the rough Heston model ``implied volatilities to parameters" map,
while Section~\ref{sec:discussion} 
develops the interpretability analysis for the latter.
We provide conclusions in Section~\ref{sec:conclusion}, as a review of the main contributions of this work and as  hints for further developments.

\section{An introduction to Interpretability analysis}
\label{sec:interpr}

The black-box feature of machine learning, 
which refers to the impossibility of mapping parameters to specific output characteristics, 
poses an issue for reliability of NN even in simple applications such as speeding up mathematical models calibrations or learning mathematical models input-output parameter maps to a satisfactory accuracy level. 
As mentioned in the introduction, the term "interpretability" lacks a precise definition, but~\cite{millerExplanation} offers a non-mathematical definition: \textit{`the degree to which a human can understand the cause of a decision'}. 
The more interpretable a machine learning model, the simpler to understand why specific judgments or predictions have been made.
When building a model, not only is its accuracy fundamental, we need also to consider how these outputs are derived from the inputs and how stable this mapping is.
Thus, interpretability can be advantageous in two scenarios:
(i) with a thorough understanding of the original mathematical model, one may test whether the learned NN map from rough Heston outputs (the model implied volatility surface)  to rough Heston inputs (the model parameters) corresponds to an intuitive understanding of the model;
(ii) with a lack of model expertise, or for mathematical models that are already little transparent to start with, one may employ interpretability methods of the learned NN map to improve the comprehension of the model.
Overall, two machine learning interpretability classifications have been devised~\cite{molnar2020interpretable}:
local and global.

\subsection{Local interpretability}
\label{local}
According to~\cite{molnar2020interpretable}, in a local interpretability model, 
each individual forecast can be explained.
Let~$f$ be the machine learning model to be interpreted and~$\widehat{f}$ the prediction given by a neural network. 
To understand the prediction $y=\widehat{f}(x)$ with regards to input~$x$, we introduce the simplified input~$x'$  of which each component is binary, 
with~$0$ when the feature value is equal to the average of the data set and~$1$ otherwise. 
For each~$x$, let further define the function~$h_x$ such that $x=h_x(x')$.

Defining local interpretability in general can be challenging. A possible, informal approach for interpretability around an input $x$ would be this. Given an input~$x$, an interpretability model is a local model~$g$ on~$x'$ such that $g(z') \approx \widehat{f}(h_x(z'))$ for $z' \approx x'$. 
A more precise definition would specify the approximations using neighborhoods: given an input~$x$, an interpretability model is a local model~$g$ on~$x'$ such that, for every $\varepsilon >0$, there exists $\delta>0$ such that $|z'-x'|\le \delta$ implies $|g(z') - \widehat{f}(h_x(z'))| \le \varepsilon$. Here $|\cdot|$ can be the standard Euclidean norm or any suitable alternative norm. 
To clarify the notations, we present a brief example. Consider a system with three input features, and the input
$x=[x_1,x_2,x_3]$. 
We can decide to eliminate some of the features (vector components) from the input, by replacing that feature with its average over the dataset. Assume that the averages of the three features~$x_1$, $x_2$ and~$x_3$ are~$m_1$, $m_2$ and~$m_3$.
If we decide to eliminate the second feature, for example, we get the vector 
$[x_1,m_2,x_3]$. If we decide to eliminate the second and third features, we get $[x_1,m_2,m_3]$. Now the notation~$x'$ will denote a vector with binary components corresponding to presence of an actual value (1) or absence of that value, replaced by  the mean (0). Absence of a feature is equivalent to that feature being replaced with its average in $x$ and with a zero in $x'$. So for example if 
$x=[x_1,x_2,x_3]$ then $x'=[1,1,1]$ because all three components are present. 
If instead $x=[x_1,m_2,x_3]$ the second component is absent, because it has been replaced with its mean, so $x'=[1,0,1]$. Finally, if $x=[x_1,m_2,m_3]$ then both the second and third feature are absent, so $x'=[1,0,0]$. When using sets of integers as indices of a vector, we mean that all the components of the vector that are not in the index have to become absent. So for example  $x_{\{2\}} =[m_1,x_2,m_3]$ and $x'_{\{2\}}=[0,1,0]$, $x_{\{1,2\}} = [x_1,x_2,m_3]$ and $x'_{\{1,2\}} = [1,1,0]$ and so on. Finally, expressions like $x'_{\{1,3\}\setminus\{3\}} = [1,0,0]$ mean that we removed feature $3$ from the set $\{1,3\}$ and so we are left only with feature 1.  
We shall consider the subclass of interpretability models~$g$ having the additive \emph{feature attribution property}, or shortly "additive"~$g$,  where the model attributes the effect~$\phi_i$ to each feature~$i$, as
$$
g(z')=\phi_0+\sum_{i} {z' _i} \phi_i.
$$

\subsection{Global interpretability}
\label{global}

A method to determine which input properties have the greatest overall impact on a model's output across all predictions is provided by global interpretability. 
The major distinction between local and global interpretability is that the global method provides a conclusion based on an average of all predictions. In other words, global interpretability is used to determine how much each input characteristic contributes to the final outcome, using all data rather than just one specific instance.
The theory of Shapley values comes from cooperative game theory and aims to assign profits and losses resulting from group activity of internal members. 
More precisely, it is computed as the average marginal contribution of a feature value across all possible coalitions;
in this game theoretic language, the game is represented by the prediction task for a single instance of the dataset, 
the P\&L is the actual prediction for this instance minus the average prediction for all instances
and the players are the feature values of the instance.
Mathematically, the problem can be set up as: assume a game with~$n$ players
$\Pb = \{p_1,\ldots, p_n\}$. 
Given a cooperation of players $G\subset \Pb$, 
$f(G)$ represents the average sum of gain/cost that members of~$G$ work together to generate. 
Shapley Values assign gains/costs to collaborative players as follows:
the attribution to the member~$p_k \in \Pb$ is
$$
\phi_{k} := \sum_{G \subseteq \Pb\setminus\{p_k\}} 
\frac{|G|! \Big(|\Pb|-|G|-1\Big)!}{|\Pb|!}\Big(f(G \cup\{p_k\}) - f(G)\Big),
$$
where $|G|$ denotes the cardinality of the set~$G$ and $|\Pb|=n$ is the cardinality of $\Pb$.
This can be interpreted as
$$
\phi_{k}=\sum_{\text{alliance without } k} \frac{\text{marginal contribution when $k$ joins the alliance}}{\text{number of possible alliances excluding $k$}}.
$$
Specifically, in the context of machine learning interpretability, the feature attribution in the presence of~$M$ features (dimension of $x$, $x'$ and $z'$) is 
$$
\phi_{k}\left(\widehat{f}, x\right)
 = \sum_{I \subseteq \{1,2,\ldots,M\}\setminus\{k\}} 
\frac{\left|I\right|! \Big(M-\left|I\right|-1\Big)!}{M !}
\left\{\widehat{f}\left(h_{x}\left(z^{\prime}_{I \cup \{k\}}\right)\right) - \widehat{f}\left(h_{x}\left(z^{\prime}_I\right)\right)\right\},
$$
where $z'_I$ is the vector $z'$ with all the components having indices that are not in $I$ set to zero, and where~$\widehat{f}$ is the prediction model. 
\texttt{SHAP} (Shapley Additive exPlanations)~\cite{lundberg2017unified} establishes a single measure of feature attribution and various practical implementation approaches, such as \texttt{Kernel SHAP}, \texttt{Tree SHAP}, and \texttt{Deep SHAP}. Additionally, \texttt{SHAP} incorporates a variety of Shapley value-based global interpretation techniques.

\section{Neural network calibration of the rough Heston model}
\label{sec:heston}

\subsection{Rough volatility (and rough Heston)}

There is ample evidence that traditional stochastic models are not able to reproduce the specificities of the volatility surfaces observed in the markets~\cite{bayer2024roughbook, bolko2023gmm}. 
Over the past few years, a new paradigm has emerged, proposing stochastic volatility models driven by a fractional Brownian motion with Hurst exponent in $(0,\half)$.
Since the seminal paper by Gatheral, Jaisson and Rosenbaum~\cite{roughvol}, where such a model was proposed based on statistical observations under the historical measure, 
a fruitful stream of research has pushed further the boundaries of such models and their 
applications~\cite{abi2019multifactor, abi2019affine, keller2018affine, el2019characteristic, euch2018perfect, guennoun2018asymptotic, jacquier2018vix}.
The overall conclusion of these papers is that, while its behaviour may be mimicked by classical Markovian models, \emph{`rough volatility as the null statistical hypothesis is very hard to beat'} (to quote Gatheral and Rosenbaum).
A fractional Brownian motion~$W^H$ is a continuous-time centered Gaussian process starting from zero, with covariance function
$$
\EE\left[W_{t}^{H} W_{s}^{H}\right]=\frac{1}{2}\left(|t|^{2 H}+|s|^{2 H}-|t-s|^{2 H}\right),
\qquad\text{for all } s,t\geq 0.
$$
Here, $H\in(0,\frac{1}{2})$ is called the Hurst parameter and describes the roughness of the paths of the process, in the sense of H\"older $(H-\eps)$ regularity for any $\eps \in (0,H)$, 
and $H=\half$ corresponds to the standard Brownian motion.
Such a process can be built on top of a two-sided standard Brownian motion~$W$ via the Mandelbrot-Van Ness representation~\cite{mandelbrot1968fractional}
\begin{equation}\label{eq:MandelbrotfBm}
W_{t}^{H} =\frac{1}{\Gamma(H+\half)}
\left(
\int_{-\infty}^{0}
\left[(t-s)^{H-\half}-(-s)^{H-\half}\right] \D W_{s}+
 \int_{0}^{t}(t-s)^{H-\half} \D W_{s}\right),
\end{equation}
almost surely for all $t\geq 0$, where~$\Gamma$ denotes the Gamma function.
Other types of fractional Brownian motions exist, and in fact the general class of interest for applications correspond to continuous Gaussian Volterra processes of the form
\begin{equation}\label{eq:fBmKernel}
W^{H}_{t} = \int_{0}^{t}K(t-s)\D W_s,
\end{equation}
almost surely for all $t\geq 0$, for some square integrable kernel function~$K(\cdot)$ on~$[0,\infty)$ and some standard Brownian motion~$W$.
In this formulation, both~$W$ and~$W^H$ generate the same filtration~\cite{decreusefond1999stochastic}
and the H\"older regularity of~$W^H$ is generated by the local singularity behaviour of~$K(\cdot)$ around the origin.
A common example uses $K(u) = u^{H-\half}\ind_{\{u>0\}}$,
giving rise to the Riemann-Liouville (or type~II) fractional Brownian motion, 
while the representation~\eqref{eq:MandelbrotfBm} can also be written as~\eqref{eq:fBmKernel}, 
albeit with a more complicated kernel function (but also with H\"older regularity $H-\eps$ for any $\eps \in (0,H)$).

Inspired by this convolutional representation, rough volatility models generally read
\begin{align*}
\frac{\D S_{t}}{S_{t}} &= \sqrt{V_{t}} \D B_{t}, \qquad S_0 >0,\\
V_{t} &= V_{0} + \int_{0}^{t}K(t-s)\left\{\left(\cdots\right) \D s + \left(\cdots\right) \D W_{s}\right\},
\end{align*}
under the risk-neutral measure (where we ignore interest rates and dividends for simplicity),
where~$S$ denotes the stock price process, $V$ the instantaneous variance process, 
and~$B$ and~$W$ are two standard Brownian motions with correlation $\rho \in [-1,1]$.
At this stage, both drift and diffusion coefficients of the variance process are left unspecified
(and of course assumptions thereupon need to be enforced to ensure weak and/or strong uniqueness).
In this project, we focus on a particular form of the variance process, namely that of the rough Heston model~\cite{el2019characteristic}
(a different formulation~\cite{guennoun2018asymptotic} is also possible, 
but the one from~\cite{el2019characteristic} is more amenable to computations and most popular in practice,
and we invite the reader to look at~\cite[Chapter~4]{bayer2024roughbook} for a full comparison),
which takes the form
$$
V_{t} = V_{0} + \frac{\kappa}{\Gamma\left(H+\frac{1}{2}\right)}
\int_{0}^{t}(t-s)^{H-\frac{1}{2}} \left\{\left(\theta-V_{s}\right) \D s
+ \nu\sqrt{V_{s}} \D W_{s}\right\},
$$
with $V_0, \kappa, \theta, \nu>0$ and $H \in (0,\half)$.
One decisive advantage of this rough Heston model is that the affine property 
of its classical counterpart is preserved (albeit in an infinite-dimensional setting),
and therefore pricing European options can be achieved via Fourier transform methods, 
as pioneered by Carr-Madan~\cite{carr1999option}, Lipton~\cite{lipton2002vol} and Lewis~\cite{lewis2001option}.
We do not repeat such an analysis here but instead refer the reader to~\cite{el2019characteristic} for full details, where a numerical scheme is provided to solve a particular form of path-dependent Riccati equation.
For path-dependent options, this approach is not amenable though, 
and one has to resort to Monte Carlo simulations. 
While this was rather intensive a few years ago, powerful new techniques have recently been devised
in this direction~\cite{bayer2020hierarchical, bayerFuk2022weak, bayer2022weak, bennedsen2017hybrid, bonesini2021functional, bonesini2023rough, fukasawa2021refinement, horvath2024functional, pannier2024path}.

\subsection{Calibration via neural networks}
\label{sec:Calibration via neural networks}
Given a set of rough Heston parameters $\pf:=\{\rho, V_0, \kappa, \theta, \nu, H\}$ living
in the parameter space $\Pp = [-1,1]\times (0,\infty)^4\times (0,1)$, 
we generate European (Call and Put) option prices and implied volatilities~$\sigimp(\cdot;\cdot, \cdot)$ over a (strike, maturity) grid $\Kk \times \Tt \subset [0,\infty)^2$.
The goal is then to construct, via a neural network to be specified, 
a map $\signn:\Theta\times\Pp\times [0,\infty)^2 \to [0,\infty)$ to learn the implied volatility for all model parameters and strikes and maturities, where~$\Theta$ is the space of neural network hyperparameters.
To do so, we define the loss function $\Ll:\Theta\to [0,\infty)$
$$
\Ll(\theta) := \sum_{(\pf,K,T)\in\Pp\times\Kk\times\Tt} \Big\| \signn(\theta;\pf;K,T) - \sigimp(\pf;K,T)\Big\|^2,
$$
and the goal is to find the minimiser
$\theta^*:=\argmin\{\Ll(\theta): \theta \in\Theta\}$.
We follow here the direct deep calibration approach from~\cite{roeder2020volatility},
where~$\signn$ is built from a feedforward neural network, similarly to~\cite{bayer2019deep}.
For each parameter set $\pf\in\Pp$ (the \emph{label}), we compute an implied volatility surface (the \emph{feature}) on the grid $\Kk\times\Tt$.
Following~\cite{brigo2021interpretability}, we in fact consider a truncated version~$\Ppb\subset\Pp$,
and generate labels~$\pf$ from this smaller set by picking them uniformly within.
The bounds used for the truncation are set in accordance with usual practice on real SPX data \cite{romer2022empirical}, as detailed in Table~\ref{tab:rHeston para}.

\begin{table}[h!]
\centering
\begin{tabular}{|c|c|c|c|c|c|c|}
\hline
   & $\rho$ & $V_0$   & $\kappa$      & $\theta$ & $\nu$ & $H$ \\ \hline
lower bound & -0.7071 & 0.0262 &   0.1206  & 0.0721 & 0.2720 & 0.1286\\ \hline
upper bound & -0.5940 & 0.0778 & 0.5041   & 0.1499 & 0.3748 & 0.1766\\ \hline
\end{tabular}
\caption{Parameter intervals for the rough Heston model.}
\label{tab:rHeston para}
\end{table}

As discussed in Section~\ref{sec:calib_result} below, 
we generated an alternative set of model parameters, where we increased the original intervals by 40\%.

\begin{table}[h!]
\centering
\begin{tabular}{|c|c|c|c|c|c|c|}
\hline
   & $\rho$ & $V_0$   & $\kappa$      & $\theta$ & $\nu$ & $H$ \\ \hline
lower bound & -1 & 0.0157 & 0.0724 & 0.0433 & 0.1632 & 0 \\ \hline
upper bound & -0.3564 & 0.1089 & 0.7057 & 0.2099 & 0.5247 & 0.4472 \\ \hline
\end{tabular}
\caption{Broader parameter intervals for the rough Heston model.}
\label{tab:rHeston para2}
\end{table}

For the out-of-sample test in Section~\ref{sec:calib_result}, we build a final data set in Table~\ref{tab:rHeston para-out}. 
The out-of-sample parameter range allows us to capture skew shapes and magnitude that are not achievable using the in-sample parameters.

\begin{table}[h!]
\centering
\begin{tabular}{|c|c|c|c|c|c|c|}
\hline
   & $\rho$ & $V_0$   & $\kappa$      & $\theta$ & $\nu$ & $H$ \\ \hline
lower bound & -0.31 & 0.12 & 0.8 & 0.24 & 0.6 & 0.47  \\ \hline
upper bound & -0.21 & 0.15 & 1.0 & 0.29 & 0.74 & 0.5 \\ \hline
\end{tabular}
\caption{Rough Heston parameters for the out-of-sample test.}
\label{tab:rHeston para-out}
\end{table}

\subsection{Description of the synthetic data set}
\label{sec:heton_calib}

We consider the following moneyness (relative to spot, centered around 1) $\times$ maturity (in years) grid
$$\Kk \times \Tt = \Big\{0.6, 0.7, 0.8, 0.9, 1, 1.1, 1.2, 1.3, 1.4\Big\} \times \Big\{0.6, 0.9, 1.2, 1.5, 1.8, 2\Big\},$$
and generate $N = 10^4$ labels $\{\pf_i\}_{i=1,\ldots,N}$ uniformly in~$\Ppb$,
yielding a family of implied volatility surfaces $\{\{\sigimp(\pf_i;K,T)\}_{K\in\Kk,T\in\Tt}\}_{i=1,\ldots,N}$,
which we split in a $85\% / 15\%$ ratio between training set and testing set.

\subsection{Neural network architectures}
\label{sec:heston_nn}

\subsubsection*{Scaling}

As explained in~\cite{brigo2021interpretability, obaid2019impact}, 
a large variance of the input data can make training challenging and often results in instability.
A common practice to avoid such discrepancies is to scale the input data beforehand;
in this spirit, we follow two procedures:
\begin{itemize}
    \item Approach 1: Scale the data to $[0,1]$;
    \item Approach 2: Scale the data to zero mean and unit variance.
\end{itemize}
We test both approaches in the following sections.
\subsubsection*{Whitening}

\begin{figure}[htbp]
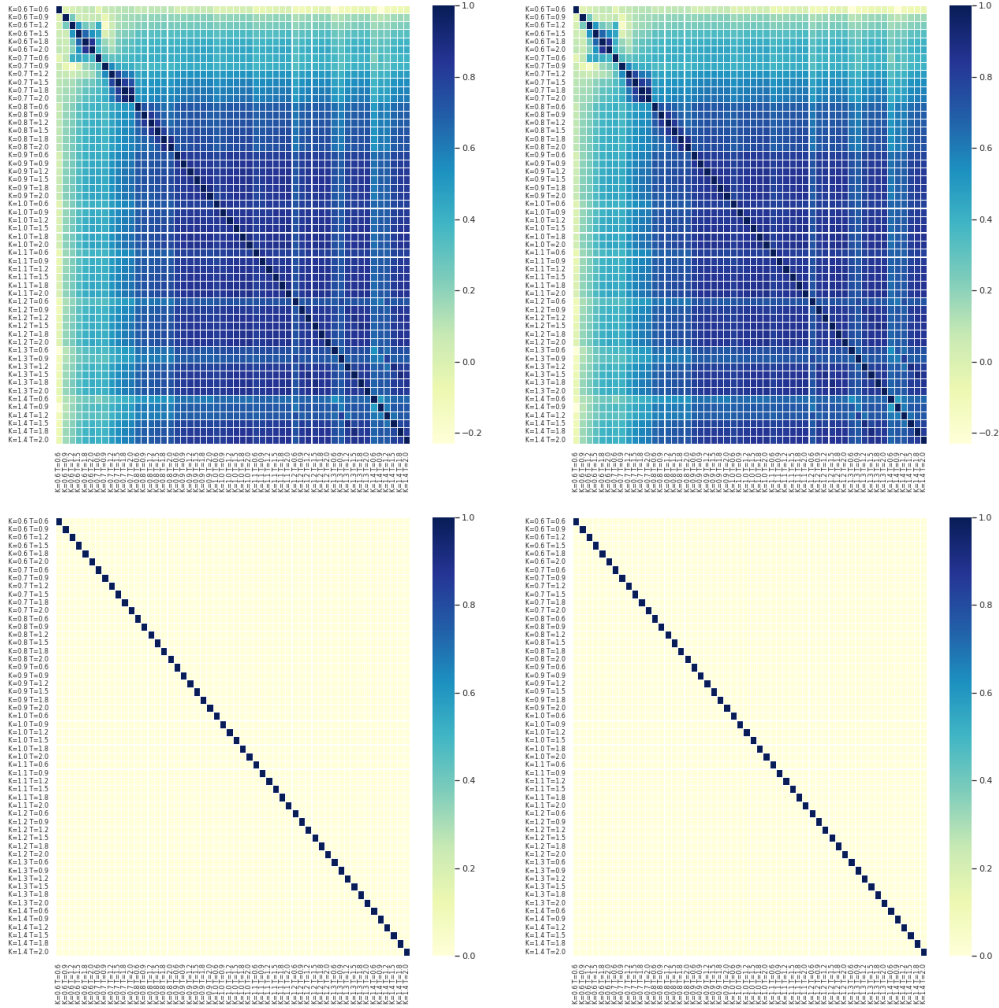

\includegraphics[width=0.485\textwidth]{corr_before1.pdf}
\includegraphics[width=0.485\textwidth]{corr_before2.pdf}\\
\includegraphics[width=0.485\textwidth]{corr_after1.pdf}
\includegraphics[width=0.485\textwidth]{corr_after2.pdf}
\caption{Correlation matrix of the volatility surface before (top) and after whitening (bottom),
using scaling Approach~1 (left) and~2 (right).}
\label{fig:corr}
\end{figure}

Before designing the neural network, it is critical to examine the scaled features and their correlations. 
The correlation matrix of the input features (the implied volatilities) is displayed in  Figure~\ref{fig:corr} highlighting the necessity of whitening.
Whitening allows one to eliminate correlation among raw data to make the input features less redundant, specifically to make the features less correlated with each other and to give all features the same variance.
Two whitening schemes are common: Principal Component Analysis (PCA) 
and Zero-phase Component Analysis (ZCA). 
They are both able to de-correlate the features and transform the correlation matrix into the identity matrix. 
We choose here ZCA as it keeps the whitened data as close as possible to the original data (in the least squares sense).
See ~\cite{brigo2021interpretability} for more details and discussion around this choice.
Given some input data~$(X_{i})_{i=1,\ldots,N}$, it works as follows: 

\begin{enumerate}[(i)]
    \item centre the data, mapping~$X$ to $\overline{X} := X - \frac{1}{N}\sum_{i=1}^{N}X_{i}$;
    
    \item compute the covariance matrix $\Sigma_X := \frac{1}{N}\overline{X}\overline{X}^{\top}$ and diagonalise it as $\Sigma_X = U\Lambda U^{\top}$ via singular value decomposition,
where~$U$ is the matrix of normalised eigenvectors and~$\Lambda$ the diagonal matrix of eigenvalues;

    \item compute the ZCA-whitening version of the data as 
    $X_w := U\Lambda^{-\frac{1}{2}}U^{\top} \overline{X}$.
\end{enumerate}

Note that the covariance matrix of~$X_w$ is precisely the identity matrix.
Figure~\ref{fig:corr} displays the correlation matrix before and after the whitening transformation. As we can see, the intra-correlation has been removed and the whitening is indeed effective. 

\subsubsection*{Feedforward neural network (FNN) architecture}
\begin{figure}[htbp]
\centering
\includegraphics[width=0.5\textwidth]{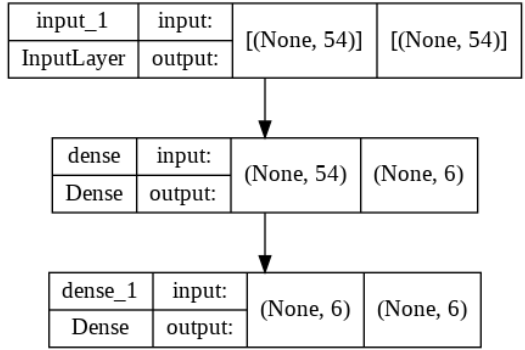}
\caption{Architecture of the fitted FNN}|
\label{fig:model}
\end{figure}

Following the pre-processing procedure, the input features (the implied volatility surfaces) are fed into a fully connected FNN. 
We proceed constructively, starting with a simple network structure and gradually increasing its complexity.
It seems that the simplest model (shown in Figure~\ref{fig:model}) with one hidden layer and a small number of neurons (with a total of~$372$ hyper-parameters) provides results sufficient for practical purposes.
We used \texttt{ELU} as activation function.
The output layer consists of the six rough Heston parameters $\pf=\{\rho, V_0, \kappa, \theta, \nu, H\}$ that need to be calibrated.
For the loss function, we use the mean-squared logarithmic function, that admits the easy interpretation as ratio of actual vs predicted values:
\begin{equation}\label{eq:LossMeanSquared}
\mathrm{Loss}\left(y,\widehat{y}\right)
:=\frac{1}{N}\sum_{i=1}^{N} \sum_{j=1}^6  
\log\left(\frac{1+y^{i}_j}{1+\widehat{y^i_j}}\right)^2,
\end{equation}
where $y=\{\rho, V_0, \kappa, \theta, \nu, H\}$ is the actual data vector and and $\widehat{y}$ is prediction output from the neural network. 
Upper indices in~$y$ denote the vector components, whereas lower indices denote the input and output index number for the neural network. 
The method is implemented using \texttt{TensorFlow} and \texttt{Keras} with
the \texttt{ADAM} optimiser.
We also use mini-batches for training, with a separation of~$20\%$ between the validation set and the training set.

\subsection{Calibration results}
\label{sec:calib_result}

\subsubsection*{Training history}

\begin{figure}[htbp]
\centering
\includegraphics[width=0.49\textwidth]{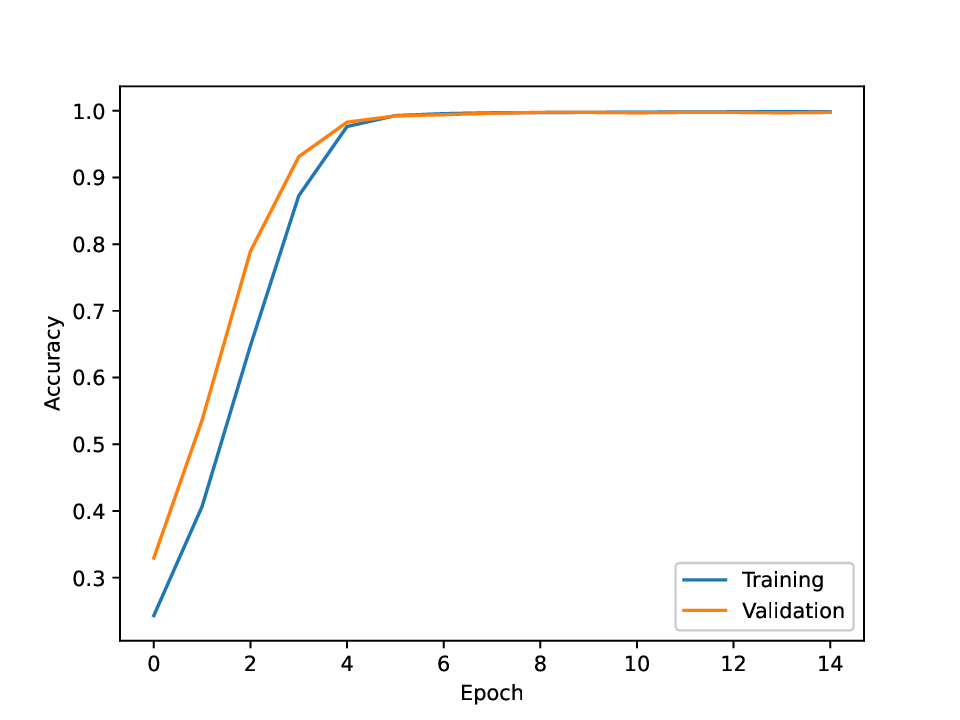}
\includegraphics[width=0.49\textwidth]{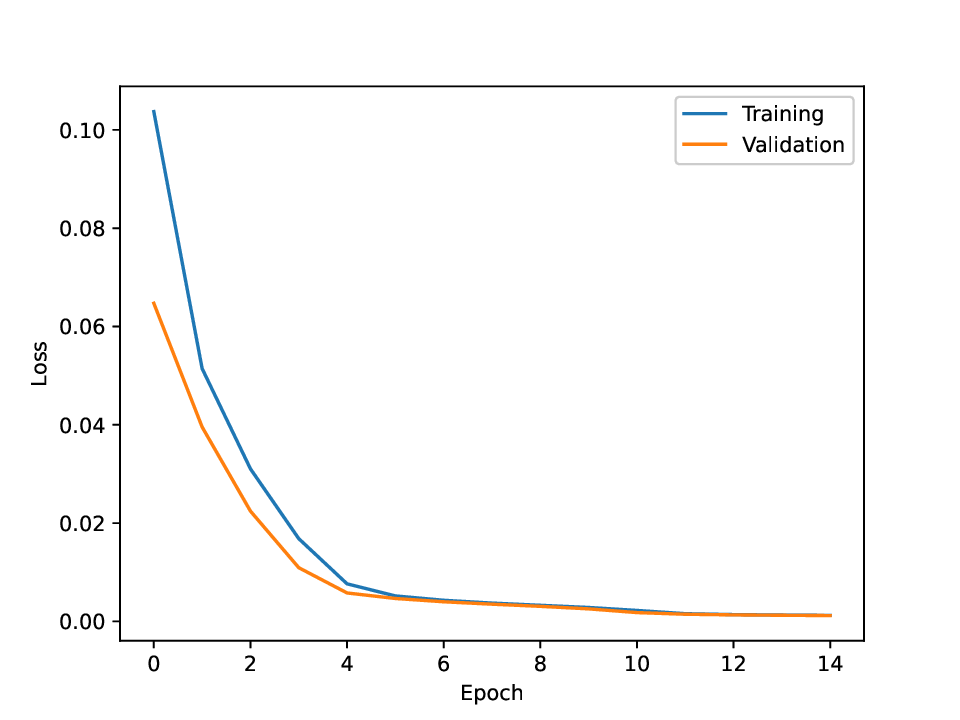}
\caption{FNN training history, with accuracy (left) and loss (right) during training.}
\label{fig:train_narrow1}
\end{figure}

To start, we employ the pre-processed data. 
Using either scaling approach, the training history appears nearly identical, and we show 
the one for Scaling approach~1 in Figure~\ref{fig:train_narrow1}.
The results suggest that the model did not overfit the training data. 
Typically, overfitting would result in the neural network's performance improving for training data but decreasing for validation data.
This does not  happen in our implementation where,  as illustrated in Figure~\ref{fig:train_narrow1}, loss and accuracy behave in the same way across training and validation data sets.
We also note that the FNN here is trained rather quickly until it reaches its limit
and the obtained level of accuracy reaches~$99\%$ for both scaling approaches.

Before adopting the chosen configuration, we experimented and repeated the FNN calibration under different configurations: we tested
mini-batch sizes of 64, 128, and 256; we tested two dropout layers with rate~0.6; finally we tested 50, 100, 150, and 200 epochs. 


\subsubsection*{Prediction errors}

The deviations between the predicted parameters and the labels for the test data are shown in Figure~\ref{fig:narrow1_test}, using again~\eqref{eq:LossMeanSquared} as error function. 
While most predictions are very accurate, there are a few outliers. 
We conjecture, similarly to~\cite{roeder2020volatility}, that this may be due to the fact that the map from parameters to data points is not bijective, and hence several different sets of parameters may yield the same data.
Table~\ref{tab:ave_narrow} shows the average prediction errors for each rough Heston parameter with the test data. 
The prediction errors for $(\nu, V_0, \theta, H$) are much lower, 
while the one for~$\kappa$ is relatively poorer than the others.
We also provide prediction errors using the training data in Table~\ref{tab:ave_narrow}. 
The fact that the neural network's performance is good not only for the training data but also for the test data
shows that training is successful in allowing the model to function effectively.  


\begin{table}[h!]
\centering
\begin{tabular}{|c|c|c|c|c|c|c|}
\hline
   & $\kappa$ & $\nu$   & $\rho$ & $V_0$      & $\theta$ & $H$ \\ \hline
training data & 0.0014             & 0.0003 & 1e-07                & 3e-05 & 0.0002                & 0.0007             \\ \hline
test data & 0.0014             & 0.0003 & 3e-08                & 3e-05 & 0.0002                & 0.0007             \\ \hline
out-of-sample data & 0.1279             & 0.0004 & 0.5593                & 0.0003 & 0.0009                & 0.0008             \\ \hline
 \end{tabular}
\caption{Prediction error mean for each parameter}
\label{tab:ave_narrow}
\end{table}

\subsubsection*{Robustness check}

To check the robustness of our fitted model, we conducted out-of-sample testing.
The average prediction error is shown in the last row of Table~\ref{tab:ave_narrow}, and the prediction errors with test data in Figure~\ref{fig:narrow1_out}. 
The forecasts for~$\kappa$ and~$\rho$ are clearly unsatisfactory, 
with average errors~$0.1279$ and~$0.5593$. 
Overall, the forecast for the other parameters appears to be accurate. 
As a result of this out-of-sample test, the FNN trained with parameters with a tighter range fails to predict properly all six parameters.


\subsubsection*{Implementation with the data set produced using the parameters with the wider range}

\begin{figure}[htbp]
\centering
\includegraphics[width=0.485\textwidth]{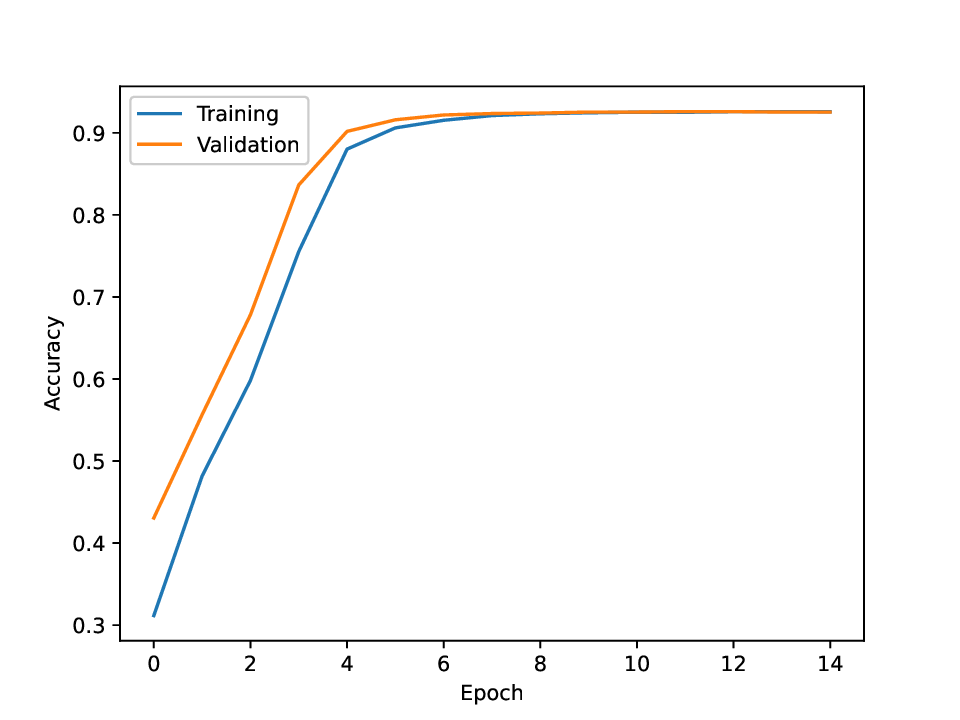}
\includegraphics[width=0.485\textwidth]{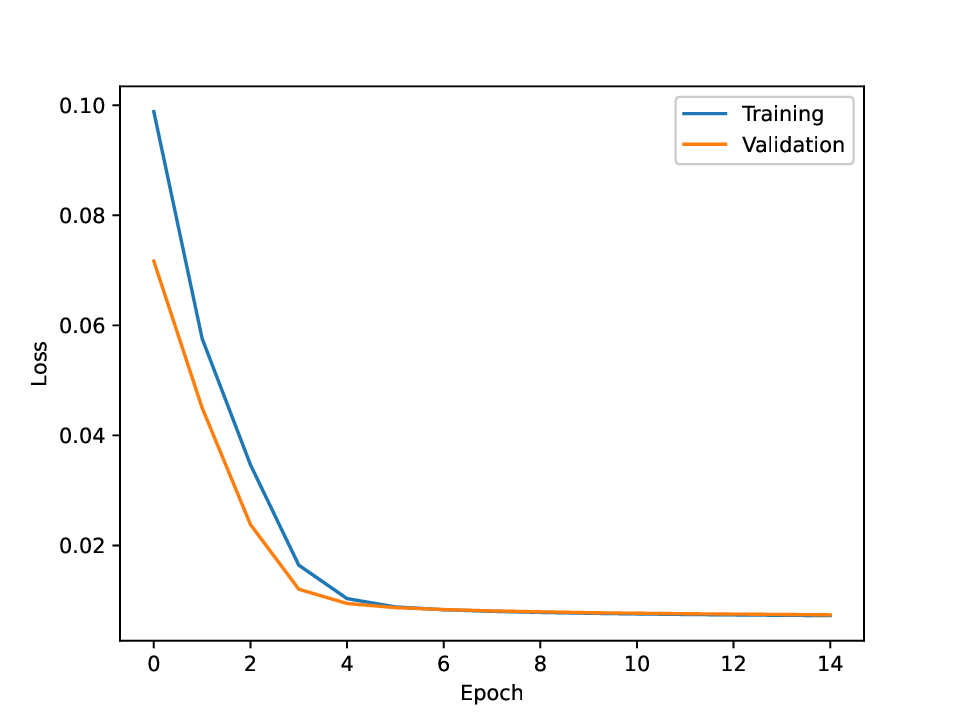}
	\caption{FNN training history with a broader data set. From the left to right, Accuracy and Loss during training, respectively.}
	\label{fig:train_wide1}
\end{figure}

As described above, we now train the FNN using the data set produced with a broader range of parameters. 
As seen in Figure~\ref{fig:train_wide1}, the training history is still relatively smooth, the learning procedure is quick to achieve stability after around 20 epochs, and the model eventually reaches an accuracy of more than $92\%$ regardless of the scaling approach.
The problem with model parameter identifiability persists, as seen in Figure~\ref{fig:wide1_test}, 
where some outliers appear to be significantly separated from the rest of the data. 
The average prediction error of the fitted FNN for the training set, validation set, 
and test data using Scaling approach~1 is once again presented in Table~\ref{tab:ave_wide} and shows that
\begin{itemize}
    \item the prediction error of FNN increases compared with that of the narrower range;
    \item the out-of-sample test indicates a higher predictive performance for~$\rho$.
\end{itemize}

We thus proceed with a FNN trained using the data set created with the parameters with the wider range for the subsequent interpretability study since it generates more robust predictions within and beyond the data.

\begin{table}[h!]
\centering
\begin{tabular}{|c|c|c|c|c|c|c|}
\hline
parameter   & $\kappa$ & $\nu$   & $\rho$ & $V_0$      & $\theta$ & $H$ \\ \hline
training data & 0.0113             & 0.005 & 0.0                & 0.0003 & 0.0008                & 0.005            \\ \hline
test data & 0.0111             & 0.005 & 0.0                & 0.0003 & 0.0008                 & 0.0055             \\ \hline
out-of-sample data & 0.1167             & 0.006 & 0.165                & 0.0012 & 0.0028                & 0.0051             \\ \hline
\end{tabular}
\caption{Prediction error mean for each parameter with a broader data set}
\label{tab:ave_wide}
\end{table}


\section{Interpretability results}
\label{sec:discussion}

\subsection{Local interpretability results}

\subsubsection*{\texttt{LIME}}

In our scenario, we should approach the NN as a regressor and use an explanatory model to achieve a local approximation around each individual prediction (here a linear model is used). 
The \texttt{LIME} library in \texttt{Python} could be used to implement this, 
with the Huber Regressor in particular selected for its tolerance to outliers. Huber optimises squared loss for the samples where $|(y - Xw - c) / \sigma| < \varepsilon$ and the absolute loss for the samples where $|(y - Xw - c) / \sigma| > \varepsilon$ (for some given tolerance~$\varepsilon$), where the model coefficients~$w$, the intercept~$c$ and the scale~$\sigma$ are parameters to be optimised.

\begin{figure}[htbp]
\centering
\includegraphics[width=0.99\textwidth]{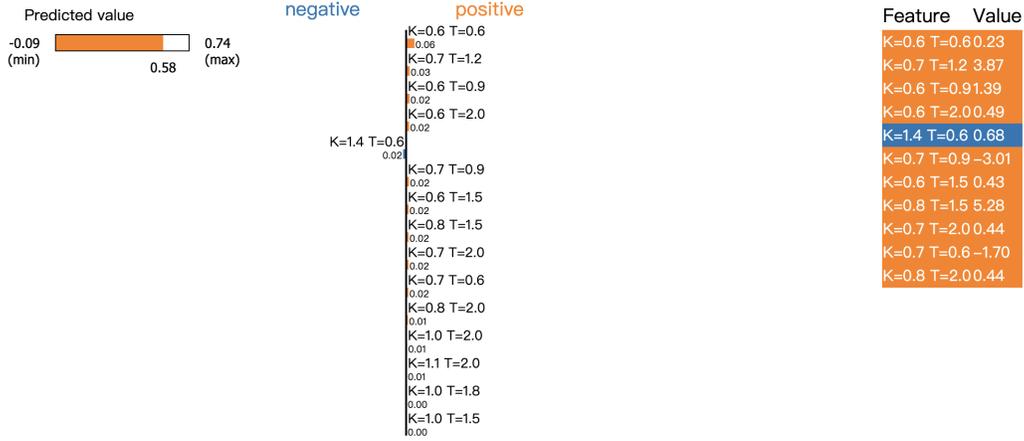}
\caption{\texttt{LIME} attribution of $\kappa$ at 0th observation (wide range of parameters)}
\label{fig:LIME_w0}
\end{figure}

\begin{figure}[htbp]
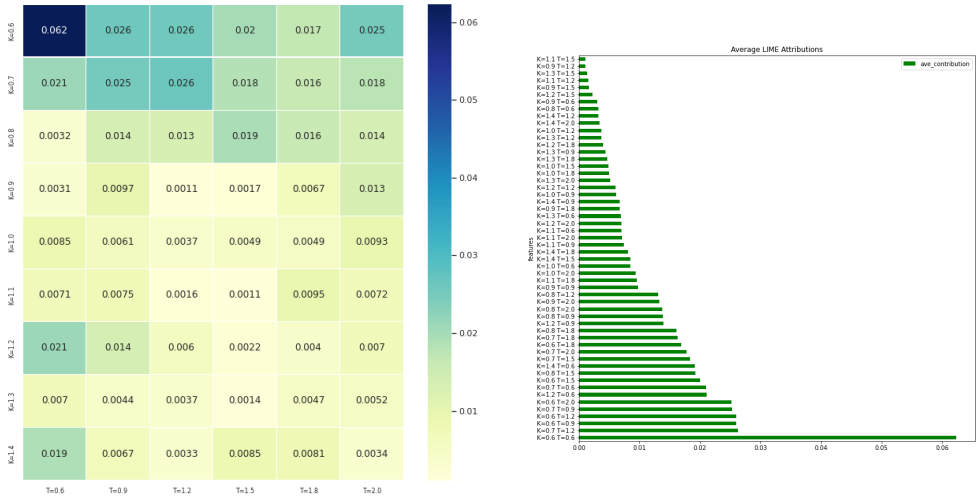

\centering
\includegraphics[width=0.485\textwidth]{heat_aveLIME_wide.pdf}
\includegraphics[width=0.485\textwidth]{aveLIME_wide.pdf}
\caption{\texttt{LIME} attributions and heat map (wide range of parameters)}
\label{fig:LIME_w}
\end{figure}

Figure~\ref{fig:LIME_w0} shows the prediction range and provides an overview of which features contribute the most and how. The orange components have a positive impact on the forecasted values, whereas the blue ones have a negative influence. The right feature-value table highlights each feature's input value with orange for positive and blue for negative. 
Figure~\ref{fig:LIME_w} looks at how each input entry impacts the model output by taking the absolute value of the contribution of each feature and calculating the average over~1500 predictions. 
Note that we only concentrate on the impact in terms of absolute values.
The light color in the heat map indicates low attribution values and the dark color high ones.

Take the first value of $\kappa$. 
As seen in Figure~\ref{fig:LIME_w0}, the prediction ranges from -0.09 to 0.74, 
with 0.58 being the first predicted value of FNN. 
The most promising feature for prediction is the short-maturity and deep in-the-money implied volatility at $(K,T) = (0.6, 0.6)$
and the next one is again deep in the money but longer maturity,  $(K,T) = (0.7, 1.2)$.
Recall that the feature values are the scaled and whitened input data with a broader range of parameters.

\bigskip

We analogously used scaled and whitened data with the narrower range of parameters to implement the \texttt{LIME} interpretability model, with similar results:
\begin{itemize}
    \item the range of the prediction covers the range of the parameter set;
    \item the deep in-the-money volatilities contribute most to the prediction;
    \item most feature contributions are positive.
\end{itemize}



\subsubsection*{\texttt{DeepLIFT}}
\begin{figure}[htbp]
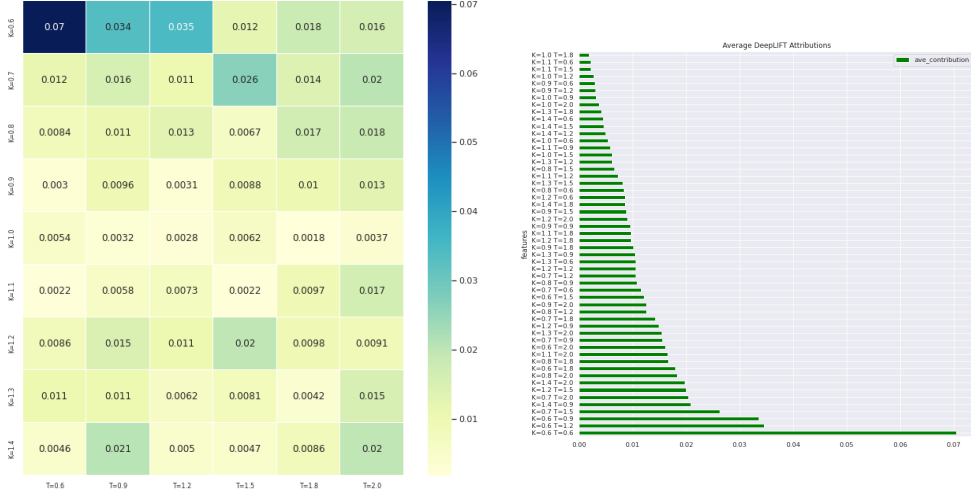

\centering
\includegraphics[width=0.485\textwidth]{heat_aveDEEPLIFT_wide.pdf}
\includegraphics[width=0.485\textwidth]{aveDEEPLIFT_wide.pdf}
\caption{Overall \texttt{DeepLIFT} attributions and heat map.}
\label{fig:DeepLIFT}
\end{figure}

\begin{figure}[htbp]
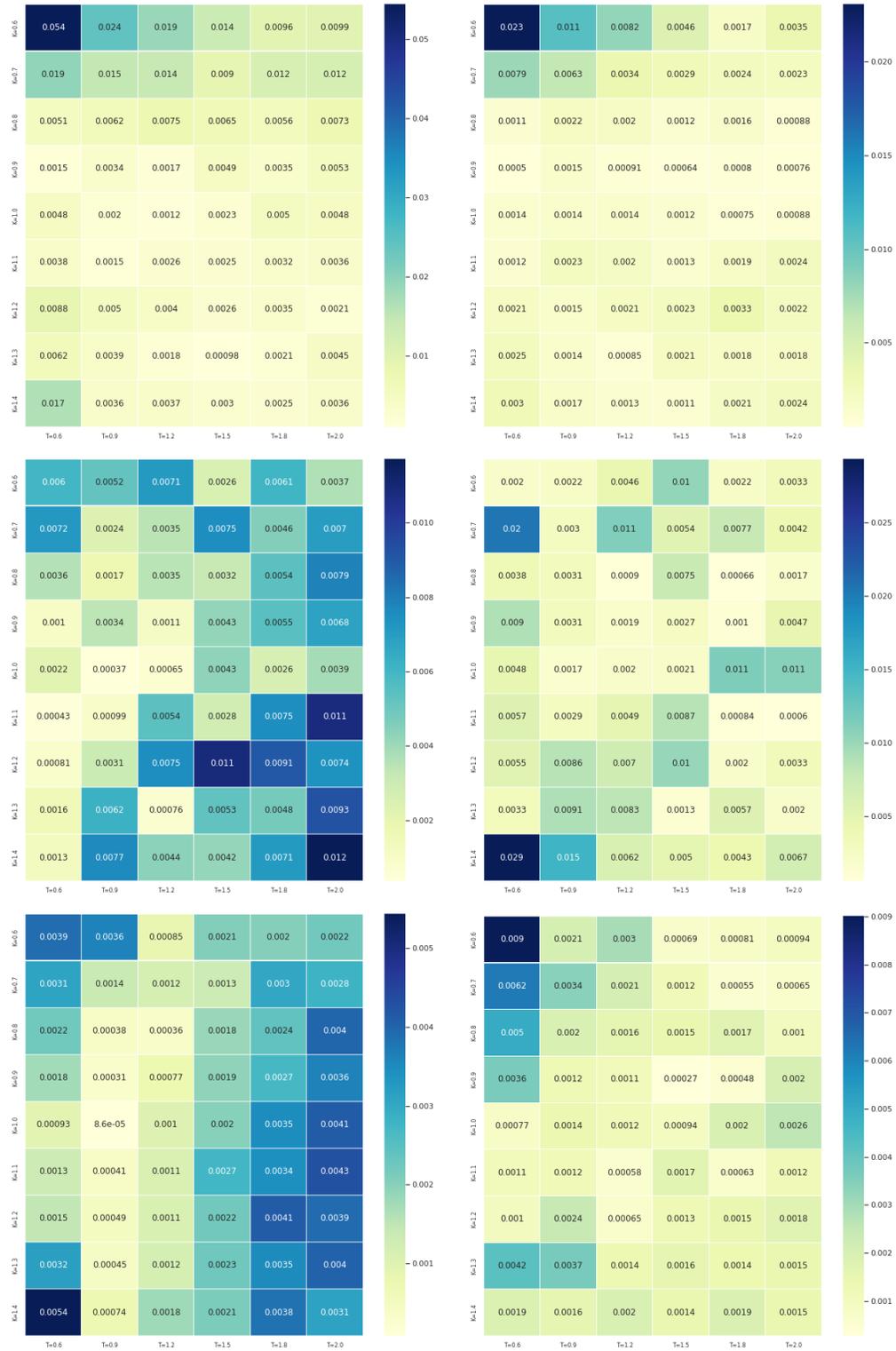

\centering
\includegraphics[width=0.485\textwidth]{heat_aveDEEPLIFT_1.pdf}
\includegraphics[width=0.485\textwidth]{heat_aveDEEPLIFT_2.pdf}
\includegraphics[width=0.485\textwidth]{heat_aveDEEPLIFT_3.pdf}
\includegraphics[width=0.485\textwidth]{heat_aveDEEPLIFT_4.pdf}
\includegraphics[width=0.485\textwidth]{heat_aveDEEPLIFT_5.pdf}
\includegraphics[width=0.485\textwidth]{heat_aveDEEPLIFT_6.pdf}
\caption{\texttt{DeepLIFT} attributions for each parameter.}
\label{fig:DEEP_para}
\end{figure}

Using the scaled and whitened test data, we create the (rescaled) \texttt{DeepLIFT} interpretability model in \texttt{Python} using the \texttt{DeepExplain} module.
Here, a zero array with the size of the input is used as the baseline by default
and the dimension of the features is $(1500, 54)$.
The all-inclusive influence on the model output $P$ as measured by 1500 test instances is shown 
in  Figure~\ref{fig:DeepLIFT}, where the deepest colours only arise around short maturities and extremely low strikes (deep in the money), in particular $(K,T) \in \{(0.6, 0.6), (0.6, 0.9), (0.6, 1.2)\}$.
The influence on each individual parameter is shown in Figure~\ref{fig:DEEP_para}:
$\kappa$ and~$\nu$ have characteristics in common with $P$ in that the primary attributions are around short maturities and extremely low strikes. 
The most significant attributions for~$\rho$ are roughly on the other side of the volatilities, 
with large maturities and extremely large strikes.
The most significant volatility for~$V_0$ appears to be distributed randomly throughout the heat map and does not follow any specific pattern.
The most significant volatility for~$\theta$ tends to have a very long maturity. 
The large attributions for~$H$ are mostly for extremely short maturities far from the at-the-money (and  mainly for deep in-the-money).

\subsubsection*{\texttt{LRP}}

Once more, we use Layer-wise Relevance Propagation \texttt{LRP} in \texttt{DeepExplain}
by propagating the prediction score backward through the network from the labels to the features.
Figures~\ref{fig:LRP} and~\ref{fig:LRP_para} suggest findings comparable to those of \texttt{DeepLIFT}:
the most important volatility for $P$, $\kappa$, and $\nu$ tends to have a very long maturity, whereas the most important volatility for~$\theta$ tends to have a short maturity and very small strikes (deep in the money). The vast majority of the attributions for the additional parameter $H$ are for very short maturities that are not in the money (mainly deep in the money). The distinction is in the attributions for $\rho$ and $V_0$: the most substantial volatility for $V_0$ emerges with short maturity on both sides of the wings, whereas $\rho$'s attributions for \texttt{LRP} are around extreme short maturity on both sides of the wings far from the ATM.

\begin{figure}[htbp]
\centering
\includegraphics[width=0.485\textwidth]{heat_aveLRP_wide.pdf}
\includegraphics[width=0.485\textwidth]{aveLRP_wide.pdf}
\caption{Overall \texttt{LRP} attributions and heat map}
\label{fig:LRP}
\end{figure}

\begin{figure}[htbp]
\centering
\includegraphics[width=0.485\textwidth]{heat_aveLRP_1.pdf}
\includegraphics[width=0.485\textwidth]{heat_aveLRP_2.pdf}
\includegraphics[width=0.485\textwidth]{heat_aveLRP_3.pdf}
\includegraphics[width=0.485\textwidth]{heat_aveLRP_4.pdf}
\includegraphics[width=0.485\textwidth]{heat_aveLRP_5.pdf}
\includegraphics[width=0.485\textwidth]{heat_aveLRP_6.pdf}
\caption{\texttt{LRP} attributions for each parameter.}
\label{fig:LRP_para}
\end{figure}

\subsection{Global interpretability results}
\begin{figure}[htbp]
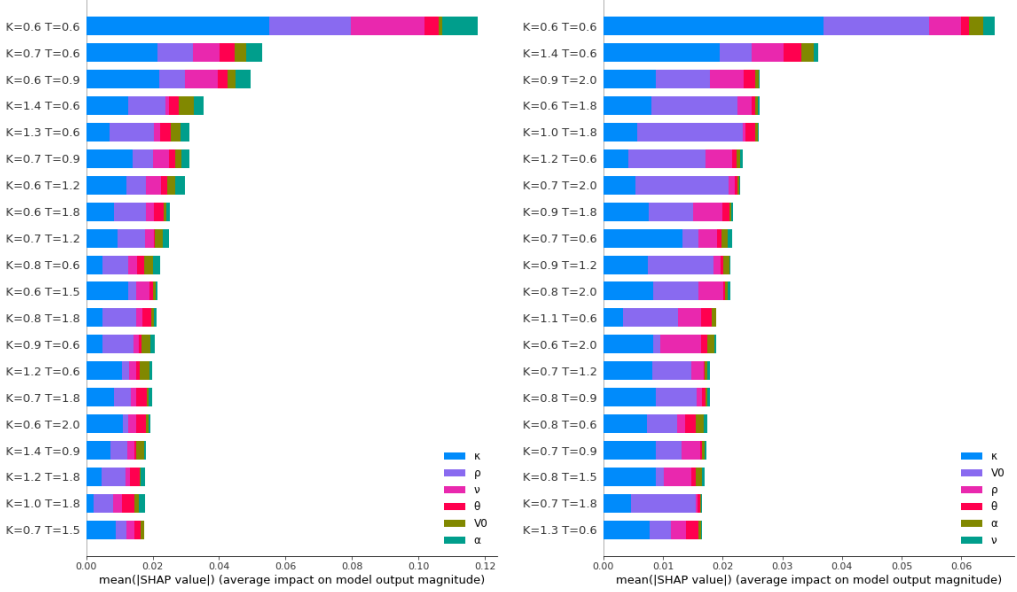

\centering
\includegraphics[width=0.485\textwidth]{shap_all_wide.pdf}
\includegraphics[width=0.485\textwidth]{shap_all_narrow.pdf}
	\caption{Overall Shapley values. From the left to right, for the larger and the smaller data set, correspondingly.}
	\label{fig:shap0}
\end{figure}

Using \texttt{DeepSHAP}, we construct a global interpretability analysis using the \texttt{Shap} module available in \texttt{Python}. As a background, we pick $1000$ samples at random from the test data set.
Figure~\ref{fig:shap0} depicts how much effect each attribute has on the overall model prediction
(the six rough Heston parameters) in the test data. 
The sum of the absolute Shapley values for each rough Heston parameter determines the order of the features. 
Here we only focus on the magnitude of the feature attribution.
As shown in Figure~\ref{fig:shap0}, we use the two data sets generated in Section~\ref{sec:Calibration via neural networks}.
For both the topmost features are $(K,T)=(0.6, 0.6)$, which illustrates that the deep in-the-money shortest-expiry volatility makes the most important contribution to the prediction, 
which is aligned with the local interpretability analysis. 
However, it should be noted that the feature influencing the prediction most is far from the money.
Furthermore, the other important features are not around the money as well. 

\begin{figure}[htbp]
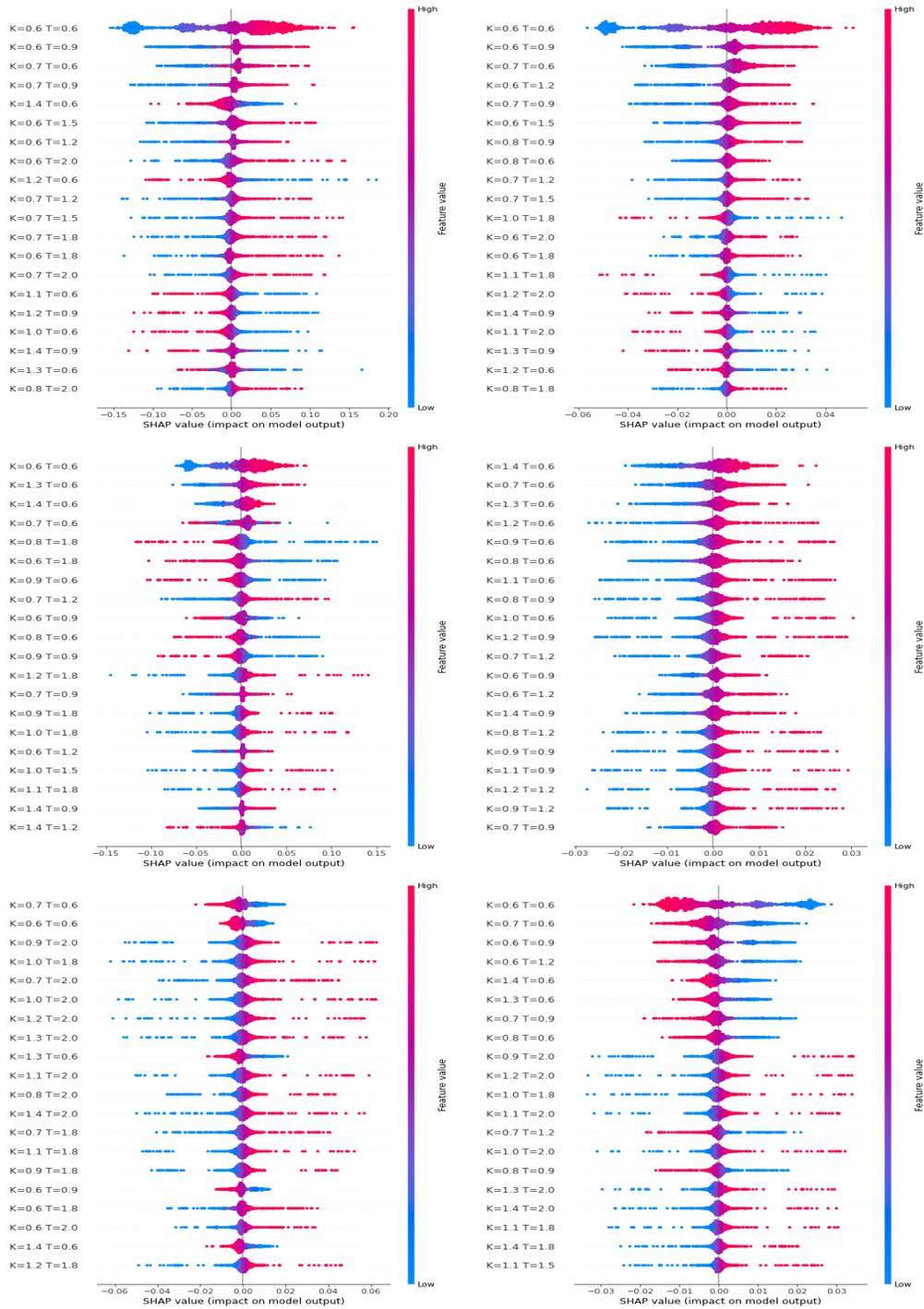

\centering
\includegraphics[width=0.485\textwidth,height=0.45\textwidth]{shap_1_wide.pdf}
\includegraphics[width=0.485\textwidth,height=0.45\textwidth]{shap_2_wide.pdf}
\includegraphics[width=0.485\textwidth,height=0.45\textwidth]{shap_3_wide.pdf}
\includegraphics[width=0.485\textwidth,height=0.45\textwidth]{shap_4_wide.pdf}
\includegraphics[width=0.485\textwidth,height=0.45\textwidth]{shap_5_wide.pdf}
\includegraphics[width=0.485\textwidth,height=0.45\textwidth]{shap_6_wide.pdf}
\caption{Shapley values for each parameter. Top down left to right: [$\kappa$, $\nu$, $\rho$, $V_0$, $\theta$, $H$]}
\label{fig:shap1}
\end{figure}

Figure~\ref{fig:shap1} gives a concise breakdown of the most crucial explanatory elements as well as how the top input features affect the label and output of the model. 
Each instance in the data collection is represented by a single dot on each feature row. 
The attribution of the feature to the prediction, which corresponds to the feature's Shapley Value, 
is shown on the x-axis.
The feature value is shown by the colour of the dots, red denoting high values and blue denoting smaller ones. 
The density is shown by the dots representing each instance piling up vertically.
The topmost features for each rough Heston parameter can be found in Figure~\ref{fig:shap1}. 
The most important features for $\kappa, \nu$ are deep in-the-money volatilities ($K \in \{0.6, 0.7\}$)
 with medium expiry.
 The features that influence~$\rho$ the most are with the extreme expiries ($T \in \{0.6, 1.8\}$)
 and either deep in the money ($K=0.6$) or deep out of the money ($K=1.4$).
The small-strike shortest-maturity volatilities ($K=0.6$) are most important to~$V_0$. 
For~$\theta$ and~$H$, the shortest-expiry deep in-the-money volatilities affect the prediction the most. 

\subsection{Discussions}
We now compare the interpretability results with the interpretation from rough Heston for each parameter. 
\begin{itemize}
\item The mean-reversion~$\kappa$ relates to small-and large-strike volatilities. 
But \texttt{LRP}, \texttt{DeepLIFT} and Shapley values all illustrate that the most influential ones are only one side (in the money) and with the short maturity. 
\item The larger the volatility of volatility~$\nu$, the more randomness in the asset price variance, which yields more obvious volatility smile features. \texttt{LRP}, \texttt{DeepLIFT} and \texttt{SHAP} provide evidence that the topmost features are deep in the money. However, further research is needed to investigate why the topmost are not on both sides and why they are mainly with short maturity, although we will provide some arguments in the comparison with the plain Heston model below.
\item The correlation~ $\rho$ describes the asymmetry of the smile. As expected, \texttt{LRP}, \texttt{DeepLIFT} and \texttt{SHAP} show that the most important features are on both sides of the wings. 
\item The initial variance~$V_0$ contains information about the short-maturity end of the smile. This is supported by \texttt{LRP} and \texttt{SHAP}.
\item The meaning of~$\theta$ as long-term average variance is supported by \texttt{LRP} and \texttt{DeepLIFT}.
But the global interpretability contradicts this by showing that the deep in-the-money features with relatively short expiry contribute most.
\item The roughness parameter~$H$ 
yields information on the wings of the smile. \texttt{LRP}, \texttt{DeepLIFT} and \texttt{SHAP} all support this interpretation and in fact, mostly with short maturities, as expected.
\end{itemize}

\section{Conclusions and further developments}
\label{sec:conclusion}
We summarise the main findings of our analysis, grouping them into the following categories.

\subsubsection*{Insights from deep calibration}

By experimenting with various data pre-processing techniques, changing the number of neurons and layers, adjusting activation functions and optimisers, and accommodating learning rate and batch size, 
we noticed that the simple one-hidden-layer architecture with ELU activation function works very well, achieving over 90\% accuracy.
We fit our feedforward neural network to two volatility smile data sets generated by the rough Heston model, one with a small range of parameters and one with a broad range.
The fitting results (and the out-of-sample data) indicate that the narrow-range FNN fails to predict all six model parameters well, and $\kappa$ and~$\rho$ in particular, while the broad-range FNN achieves good prediction.

\subsubsection*{Findings from Interpretability Models}

We focus mostly on local surrogate models (\texttt{LIME}, \texttt{DeepLIFT}, and \texttt{LRP}). 
We also give some preliminary results within the global interpretability analysis, specifically with \texttt{SHAP}. 
Overall, we find that the interpretability of the input-output map given by local methods is similar and roughly consistent with the interpretability obtained with the global method.
In fact, we may anticipate ATM volatility to be the most significant feature for calibration because for real market data it has high liquidity, and in a way it represents a sort of key point of the smile even for the synthetic type of data we use.
However, our analysis shows that 
short-maturity deep in-the-money volatilities actually provide the largest  contribution overall. This is actually in line with research on the model itself and is due to its specific behaviour driven by its rough nature \cite{kellermajid}, as we explain in the comparison with the standard Heston below. 

Furthermore
\begin{itemize}
\item the topmost feature of the correlation are on both wings of the volatility. 
\item \texttt{LRP} and \texttt{SHAP} show that short-maturity volatilities  contribute the most to~$V_0$.
\item Both \texttt{LRP} and \texttt{DeepLIFT} show that~$\theta$ is mostly driven by long-expiration volatility.
\item The Hurst parameter, the mean-reversion speed and the volatility of volatility
seem to be captured only on one side of the volatility smile, and concentrate on short maturities.
\end{itemize}
Note that the last item is not necessarily surprising, and is actually consistent with the fact that the Hurst parameter drives the short-maturity at-the-money skew of the smile, related to power-law decay  
(we refer the interested reader to~\cite{Alos07, fukasawa2017short} and~\cite[Chapter~8]{bayer2024roughbook} for precise details about this).

\subsubsection*{Comparison with the plain Heston model interpretability \cite{brigo2021interpretability}} 
We can summarize the comparison between the two models in Table~\ref{table:comparison}, which compares the heat maps readings of the different interpretability methods for the two models.

\begin{table}[h!]
\begin{tabular}{|p{2cm}|p{5cm}|p{5cm}|}
\hline 
Approach & Heston  & rHeston \\ \hline
LIME (loc) & $(K,T)=(1.1,1.8),(1.3,0.9)$ are the most significant features. but otherwise no clear pattern observed & $(K,T)=(0.6,0.6)$ is the most significant feature. Most top features are in the money (ITM) $(K=0.6/0.7)$. Small $T$ matters but not as strongly as $K$. \\
DeepLift (loc) & All significant features correspond to T=0.1/0.3, indicating short maturity. No clear pattern regarding moneyness. & $(K,T)=(0.6,0.6/0.9/1.2)$ are the most significant features. Most top features are deep ITM with short maturity. \\
LRP (loc) & All significant features correspond to $T=0.1/0.3$ and are far from at the money (ATM). &  
$(K,T) =(0.6/0.7,0.6/0.9)$  are the most significant features. Most top features are deep ITM with short maturity.
\\
SHAP (glob) &  Most significant features correspond to $T=0.1/0.3$. No clear pattern regarding moneyness. & $(K,T)=(0.6,0.6)$ is the most significant feature.  \\
Local comparison & All loc methods agree on short maturity being an important feature, except LIME & Consistency. Most important features are ITM, small maturity (almost always). \\ 
loc vs glob & Loc agree with glob on short maturities being important, except LIME & 
Consistency loc/glob. Most important features are ITM, small maturity (almost always)\\
\hline
\end{tabular}
\caption{Heat map reading of different FNN  interpretability models: Heston vs rHeston}\label{table:comparison}
\end{table}

The table gives circumstantial evidence of a sensible interpretability pattern for rough Heston. Indeed, in \cite{kellermajid} it is shown that for small maturities, the slope of left-wing implied volatility in the rough Heston model is dramatically steeper than in the non-rough Heston model. This complements known results on small-time behaviour of the at-the-money skew in rough models, which explodes at the same rate  \cite{fukasawa2017short}. This may explain why while in the standard Heston model the interpretability focuses mostly on short maturities but does not single out any strong pattern in the option moneyness, having the smile a milder and more regular pattern and being more evenly distributed, in the rough Heston model interpretability picks as most important the parts of the smile that are characteristic of rough Heston, namely the left wing. Indeed, the left wing of the smile corresponds to deep in the money options, which are the part of the smile that seems to matter most in determining the parameters through the FNN both according to local and global interpretability methods. Also, the fact that the interpretability focus is mostly on small maturities is confirmed by rough Heston exhibiting this extremely distinctive left wing behaviour for small maturities.   

We further compare, for the global method SHAP, which was the main focus in \cite{brigo2021interpretability}, the different output  parameters results in Table~\ref{Table:CompareParameters}.

\begin{table}[h!]
\begin{tabular}{|p{2cm}|p{5cm}|p{5cm}|}
\hline 
Parameter & Heston  & rHeston \\ \hline
$V_0$ & Smile wings, short maturities & Smile wings, short maturities \\
$\theta$ & Large maturities & No dominant pattern \\
$\kappa$ & Close to smile wings and both short and long maturities &  Wings of the smile, short maturities  \\  
$\nu$ & Smile convexity & Wings of the smile (one particular side), short maturities  \\
$\rho$ & Wings of the smile, short maturities & Wings of the smile, short maturities\\ 
$H$ &  NA &  Wings of the smile\\
\hline
\end{tabular}
\caption{Different parameters analysis for FNN interpretability in SHAP: Heston vs rHeston}\label{Table:CompareParameters}
\end{table}

\subsubsection*{Further developments}

We trained our neural networks on synthetic data, generated from the rough Heston model.
The next step will be to check on real, obviously noisier, data.
We did not include this analysis here though, as it would have made it hard to disentangle the interpretability of the network for the model and the fit of the model to the real data.
More generally, due to backpropagation, neural networks are non-linear and non-local. 
Local surrogate models are based on localisation and, in certain cases, linearisation, so that the surrogate interpretability model changes around each individual prediction where it is localised and possibly linearised, whereas Shapley values provide a global interpretability method that is not re-defined around each specific input. 
We hope that further research into global interpretability will shed light on the differences we highlighted between the intuitive, mathematical-finance based interpretation of the smile-parameters map and the interpretation suggested by the deep learning global interpretability method.

\bibliographystyle{siam}
\bibliography{biblio}

\newpage
\appendix
\section{Figures on calibrations via neural networks}
\vspace{-\baselineskip}

\begin{figure}[htbp]
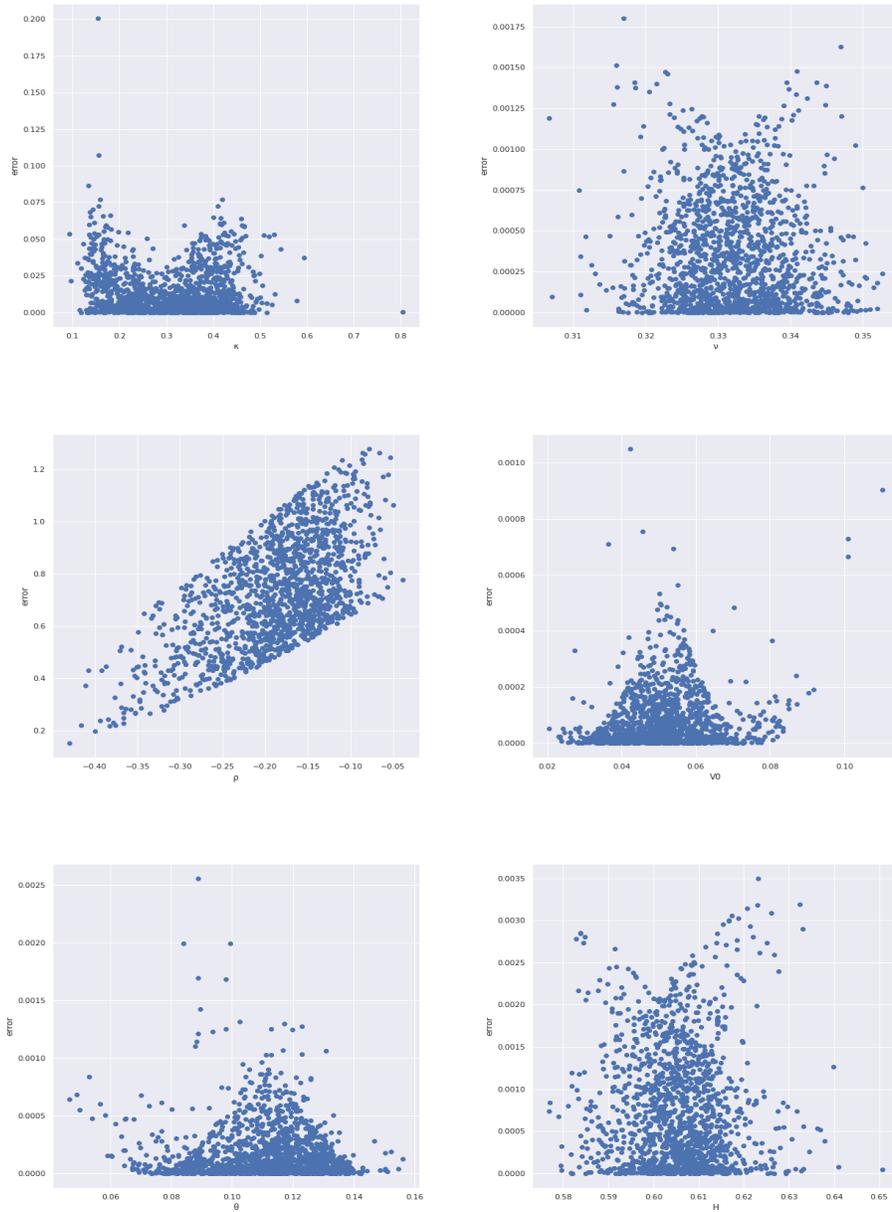

\centering
\includegraphics[width=0.45\textwidth, height=0.28\textheight]{ntest_error_k.pdf}
\includegraphics[width=0.45\textwidth, height=0.28\textheight]{ntest_error_v.pdf}
\includegraphics[width=0.45\textwidth, height=0.28\textheight]{ntest_error_rho.pdf}
\includegraphics[width=0.45\textwidth, height=0.28\textheight]{ntest_error_V0.pdf}
\includegraphics[width=0.45\textwidth, height=0.28\textheight]{ntest_error_theta.pdf}
\includegraphics[width=0.45\textwidth, height=0.28\textheight]{ntest_error_H.pdf}
\caption{Prediction errors of each parameter with the test data. From the left to right, Prediction error for $\kappa$, $\nu$, $\rho$, $V_0$, $\theta$, $H$, respectively.}
\label{fig:narrow1_test}
\end{figure}

\begin{figure}[htbp]
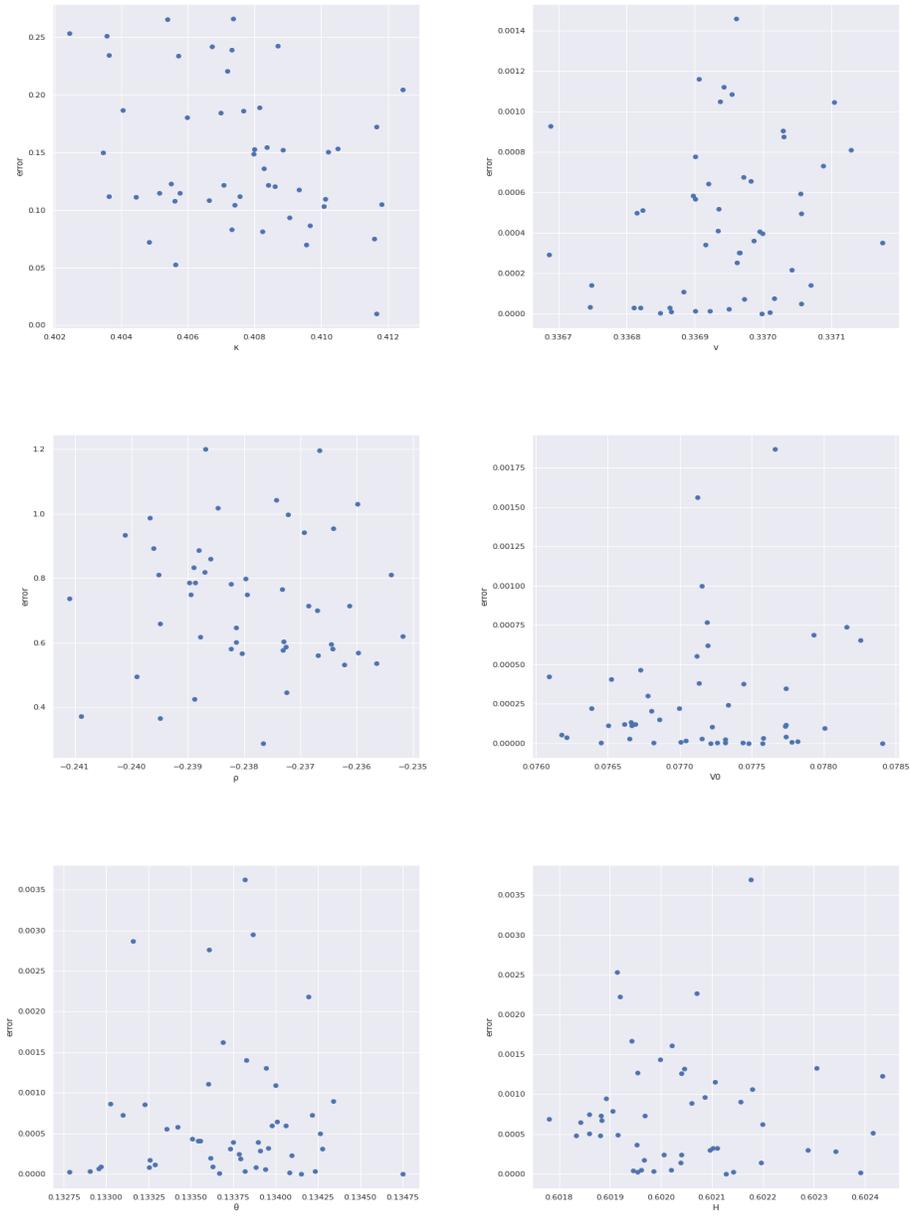

	\centering
\includegraphics[width=0.45\textwidth, height=0.28\textheight]{narrow_out_k.pdf}
\includegraphics[width=0.45\textwidth, height=0.28\textheight]{narrow_out_v.pdf}
\includegraphics[width=0.45\textwidth, height=0.28\textheight]{narrow_out_rho.pdf}
\includegraphics[width=0.45\textwidth, height=0.28\textheight]{narrow_out_V0.pdf}
\includegraphics[width=0.45\textwidth, height=0.28\textheight]{narrow_out_theta.pdf}
\includegraphics[width=0.45\textwidth, height=0.28\textheight]{narrow_out_H.pdf}
\caption{Prediction errors of each parameter with the out-of-sample data.}
\label{fig:narrow1_out}
\end{figure}

\begin{figure}[htbp]
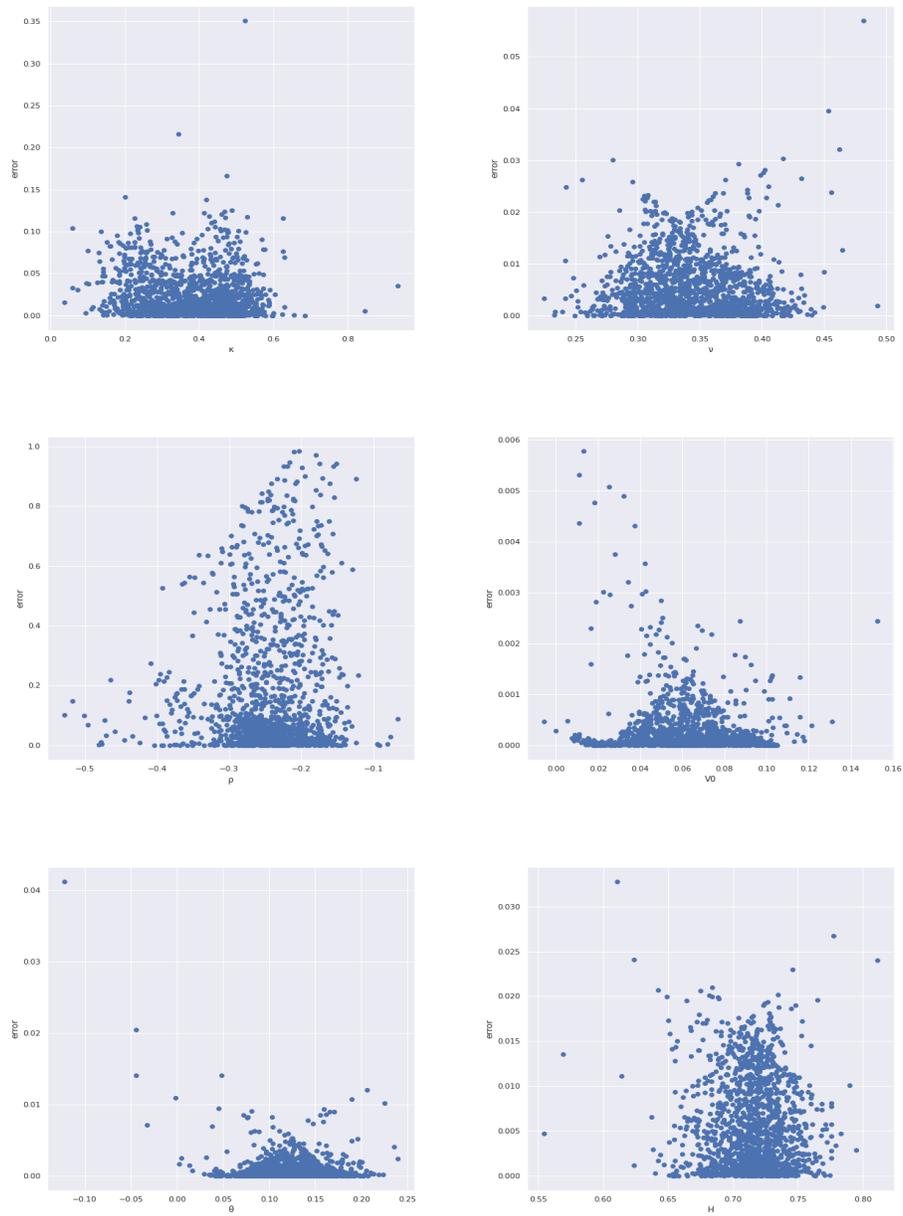

	\centering
\includegraphics[width=0.45\textwidth, height=0.28\textheight]{wtest_error_k.pdf}
\includegraphics[width=0.45\textwidth, height=0.28\textheight]{wtest_error_v.pdf}
\includegraphics[width=0.45\textwidth, height=0.28\textheight]{wtest_error_rho.pdf}
\includegraphics[width=0.45\textwidth, height=0.28\textheight]{wtest_error_V0.pdf}
\includegraphics[width=0.45\textwidth, height=0.28\textheight]{wtest_error_theta.pdf}
\includegraphics[width=0.45\textwidth, height=0.28\textheight]{wtest_error_H.pdf}
\caption{Prediction errors of each parameter with the broader data set.}
\label{fig:wide1_test}
\end{figure}

\clearpage

\section{Figures on local interpretability analysis}

\begin{figure}[htbp]
\centering
\includegraphics[width=0.99\textwidth]{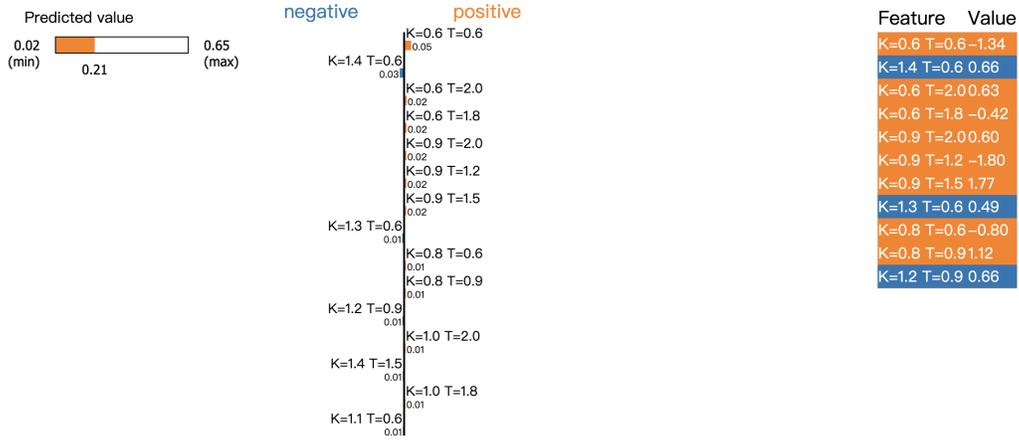}
\caption{\texttt{LIME} attribution of $\kappa$ at 0th observation (narrow range of parameters)}
\label{fig:LIME_n0}
\end{figure}

\begin{figure}[htbp]
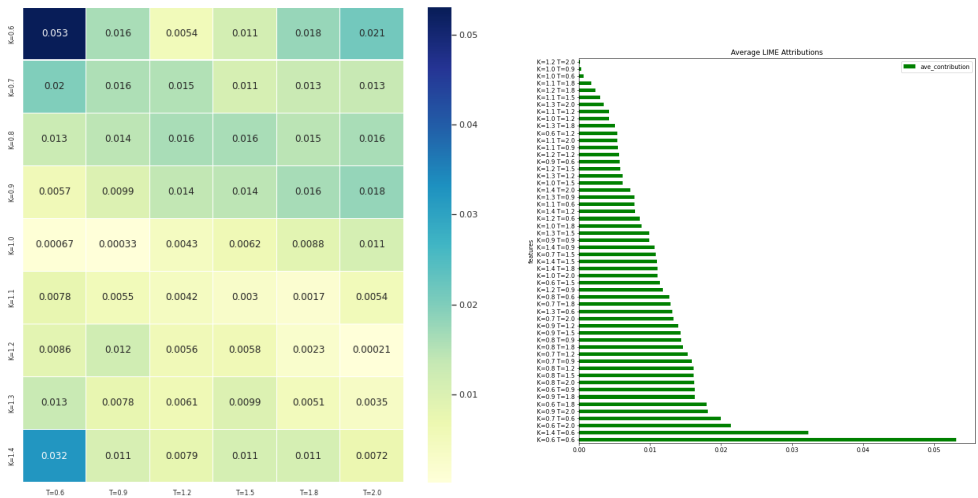

\centering
\includegraphics[width=0.485\textwidth]{heat_aveLIME_narrow.pdf}
\includegraphics[width=0.485\textwidth]{aveLIME_narrow.pdf}
\caption{\texttt{LIME} attributions and heat map (narrow range of parameters)}
\label{fig:LIME_n}
\end{figure}

\newpage

\end{document}